# Harmonic mean, the Gamma factor and Speed of Light

Chandru Iyer, Techink Industries, C-42 Phase II, Noida, India

chandru_i@yahoo.com

Abstract: The relationship between the harmonic mean and special relativity is concisely elucidated. The arguments in favor and against SRT are explored. It is shown that the ratio of the speed of light to the harmonic mean of the onward and return speeds of light in a moving frame under Newtonian mechanics, when equitably distributed between space and time as a 'correction', leads to the Lorentz transformation. This correction implies an 'apparent' contraction of objects and time dilation. However, the symmetry of the onward and inverse transformations give a different meaning to the gamma factor.

There are various ways of determining averages. We are familiar with terms such as arithmetic mean, geometric mean and harmonic mean. Which of this is appropriate in a given situation depends on the nature of the processes involved. If one travels at different speeds for equal time intervals then the average speed will be the arithmetic mean. If one travels equal distances at different speeds, then the average speed will be the harmonic mean (of the different speeds).

If a light ray is traveling at a speed of c, then for an object moving at *v*, according to Newtonian physics, this light ray should travel at c-*v* in one direction and c+*v* in another direction. If the light ray travels a distance s, gets reflected and returns, the average speed will be the harmonic mean, under Galilean-Newtonian mechanics.[1]

$$\frac{2s}{\frac{s}{c-v} + \frac{s}{c+v}} = \frac{(c^2 - v^2)}{c} = c[1 - (v^2/c^2)]$$

Thus, the average speed of light for the moving frame should have been $c[1-(v^2/c^2)]$ as envisaged by the stationary frame. Here, we immediately find the famous gamma factor of relativity appearing from a simple application of the harmonic mean.

However, this average speed of light is Newtonian and under relativity this has to be equal to c. Therefore, the factor $[1-(v^2/c^2)]$ needs to be neutralized. This is easily achieved under relativity by contracting the moving rulers by the factor $(1/\gamma) = \sqrt{1 - v^2/c^2}$ whence the distance and speed increase by the factor γ and further making the moving clocks run slow by the same factor 1/γ, which once again increases the speed by the factor γ.

---

[1] The light ray is traveling at speed c, isotropically, with respect to inertial frame K. The object and mirror are co-moving with inertial frame K' at speed v in the positive x direction. The light ray starts at the object and returns to the object after reflecting from the mirror. The distance between the object and mirror remains constant always and is observed to be s by K.

So what stands between the arithmetic and harmonic means of (c+*v*) and (c-*v*) is the factor [1-(*v*²/c²)]. This factor was distributed equally between space and time each carrying a 'correction' of $\sqrt{1-v^2/c^2}$. The factor thus stands neutralized and the arithmetic mean becomes equal to the harmonic mean.

The following table records the observations of the stationary frame, and the moving frame. The observations of the stationary frame are the same under Newtonian and Relativistic mechanics. The moving frame's observations differ under Newtonian and Relativistic mechanics.

Table: <u>Round Trip of Light Ray in a **'Moving Frame'**</u>

| SL_NO | | 'Stationary Frame' (Newtonian as well as Relativistic Observations) | 'Moving Frame' (Newtonian Observations) | 'Moving Frame' (Relativistic Observations | Remarks |
|---|---|---|---|---|---|
| 1 | Onward Journey Distance | sc/(c-v)  (Note below Table) | s | sγ | |
| 2 | Onward Journey Time | s/(c-v) | s/(c-v) | sγ/c | Source of Asynchron--ization |
| 3 | Onward Journey speed | c | (c-v) | c | |
| 4 | Return Journey Distance | sc/(c+v) | s | sγ | |
| 5 | Return Journey Time | s/(c+v) | s/(c+v) | sγ/c | Source of Asynchron--ization |
| 6 | Return Journey Speed | c | (c+v) | c | |
| 7 | Total Distance | 2s / [1-($v^2/c^2$}] | 2s | 2sγ | |
| 8 | Total Time | 2s / c[1-($v^2/c^2$}] | 2s / c[1-($v^2/c^2$}] | 2sγ/c | |
| 9 | Average Speed | c | c[1-($v^2/c^2$}] = c/γ² | c | γ² is neutralized as discussed |

Note: The Mirror is in the moving frame. During the time interval the light ray travels a distance of s, the mirror moves further by sv/c; the infinite converging geometric series thus obtained gives the distance traveled by the light ray to reach the mirror as sc/(c-v)

The onward journey distance traveled by the light ray (as observed by K) may also be calculated by assuming it as 'd'. The time taken for the onward journey is d/c. In this time the mirror moves an additional distance of dv/c. Thus d=s+(dv/c) or d=sc/(c-v). However, the onward journey distance traveled by the light ray as observed by K' remains s.

From the above table, one can see the difference in the observation of the elapsed time during the onward and return journeys. The Newtonian physicists expect the onward journey elapsed time to be s/(c-$v$); the relativists expect the same to be s/(c-$v$) by the stationary observers and (s$\gamma$/c) by the moving observers. This difference leads to the observed asynchronization of spatially separated clocks of K' as observed by K and vice-versa.. Similarly, the elapsed time for the return journey is expected to be s/(c+$v$) by the Newtonian physicists; the relativists expect the same to be s/(c+$v$) by the stationary observers and (s$\gamma$/c) by the moving observers. <u>The geometrically equitable distribution of the factor [1-($v^2/c^2$)] between space and time ensures an internally consistent system.</u>

However, the intriguing aspect of relativity arises when it claims that the rulers of the moving object contracted only as observed by the stationary observer and in fact the rulers of the stationary observer were also observed to contract as observed by the moving observer. At this point the theory assumes a metaphysical perplexity. It also claims that the rods did not actually contract but were only observed to have contracted. The same arguments apply to the running of clocks. The clocks did not run slower actually but were only observed to be running slow.

The objections to the theory are based on the premise, "if what we observe has not actually happened then our observations need to be reviewed and appropriately corrected". The possibility that two rods can contract as observed by each other and we can never determine what actually happened in an absolute sense, devoid of the association of ourselves as co-moving with any one of the rods, has been exercising the minds of scientists over a century now

The above analysis and arguments clearly bring out the point that the constancy of the speed of light is possible when the moving rods contract and moving clocks run slow by a factor (1/$\gamma$) = $\sqrt{1-v^2/c^2}$ . When this contraction and slow running are mutual, they become only an apparition and not an actual happening. When these physical changes do not actually happen, then the arguments fail. When we cannot explain phenomena by a concrete set of principles, but only principles that are valid when associated with particular observers, then it becomes difficult to understand the validity of these principles.

Many physicists [1,2,3,4,5,6,7] have been voicing these concerns for a century now. I recently developed the above derivation of the relationship between the harmonic mean and the gamma factor and was very surprised that how in just a few steps, the relativistic length contraction and time dilation can be derived with ease. However, for my arguments to hold, these effects should be actual and not apparent. I have also addressed the question "can symmetry imply equivalence?" in [8]

If the cause of the contraction of the rulers and slower running of clocks is the movement of the object, then this cause clearly distinguishes between what is moving and what is stationary. Further it implicitly assumes that the stationary rods did not contract but only the moving rods.

Therefore, if the moving rulers did not contract, then the constancy of the speed of light is difficult to explain by logic. If they did contract, then there is a clear distinction between the moving ruler and the stationary ruler. This violates the first postulate of special relativity, which essentially states that all bodies in uniform relative motion are equivalent.